\documentclass[11pt,letterpaper]{article}
\pdfoutput=1

\usepackage[includeheadfoot,
            marginratio={1:1,2:3}, 
            width=412pt, 
            height=688pt,]{geometry}
\usepackage{amsmath}
\usepackage{amsfonts}
\usepackage{amssymb}
\usepackage{stmaryrd}
\usepackage{cite}
\usepackage{empheq}
\usepackage[arrow,matrix,curve]{xy}

\usepackage{filecontents}
\usepackage[hidelinks]{hyperref}
\usepackage{graphicx} 
\usepackage{pstricks}
\usepackage[utf8]{inputenc}

\newcommand{\nc}{\newcommand}
\nc{\beq}{\begin{equation}}
\nc{\eeq}{\end{equation}}
\nc{\bea}{\begin{eqnarray}}
\nc{\eea}{\end{eqnarray}}
\def\ov{\overline}

\newcommand{\eq}[1]{\begin{equation}
                     \begin{split} #1 \end{split}
                     \end{equation}}

\begin{document}

%%%%%%%%%%%%%%%%%%%%%%%%%%%%%%%%%%%%%%%%%%%%%%%
%%%%%%%%%%%%%%%%%%%%%%%%%%%%%%%%%%%%%%%%%%%%%%%

\vspace*{-1.5cm}
\begin{flushright}
  {\small
  MPP-2019-204\\
  }
\end{flushright}

\vspace{2.5cm}
\begin{center}
{\LARGE
Quantum Log-Corrections to Swampland Conjectures\\[0.2cm]
} 
\vspace{0.6cm}

\end{center}

\vspace{0.5cm}
\begin{center}
  Ralph Blumenhagen, Max Brinkmann, Andriana Makridou
\end{center}

\vspace{0.1cm}
\begin{center} 
\emph{Max-Planck-Institut f\"ur Physik (Werner-Heisenberg-Institut), \\ 
   F\"ohringer Ring 6,  80805 M\"unchen, Germany }

\vspace{0.2cm}

 \vspace{0.3cm} 
\end{center} 

\vspace{0.79cm}
%%%%%%%%%%%%%%%%%%%%%%%%%%%%%%%%%%%%%%%%%%%%%%%

\begin{abstract}
Taking the anti-de Sitter minimum of KKLT and the large volume
scenario at face value, we argue for the existence of logarithmic 
quantum corrections to AdS swampland conjectures. 
If these conjectures receive such corrections, it is natural to suspect
that they also arise for other swampland conjectures,
in particular  the dS swampland conjecture.
We point out that the proposed $\log$-corrections are 
in accord with the implications of the recently proposed
trans-Planckian censorship conjecture. We also comment
on the emergence proposal in the context of both perturbative flux models
and the  KKLT construction.
\end{abstract}

%%%%%%%%%%%%%%%%%%%%%%%%%%%%%%%%%%%%%%%%%%%%%%%

\clearpage

\tableofcontents

\section{Introduction}
\label{sec:intro}

String theory and in particular its application to low energy physics
is  experiencing a period of new insights that question older
arguments based on the landscape picture.
The swampland program
\cite{Vafa:2005ui,ArkaniHamed:2006dz,Ooguri:2006in} 
intends to extract  a set of relatively simple quantitative features
that  low-energy effective field
theories should satisfy in order to admit a  UV completion
to a consistent theory of quantum gravity (see \cite{Palti:2019pca} for a recent review). 
By now several swampland conjectures have been proposed
{\cite{Klaewer:2016kiy,Ooguri:2016pdq,Palti:2017elp,Obied:2018sgi,Andriot:2018wzk, 
Cecotti:2018ufg,Garg:2018reu,Ooguri:2018wrx,Klaewer:2018yxi,Heckman:2019bzm,Lust:2019zwm,Bedroya:2019snp,Kehagias:2019akr}},
which  also induced  further
new developments like  the  emergence
\cite{Heidenreich:2017sim,Grimm:2018ohb,Heidenreich:2018kpg} of infinite
distances in field space or the appearance of towers of light strings \cite{Lee:2018urn,Lee:2018spm,Lee:2019wij}.

In this paper we focus on two such conjectures dealing with AdS vacua. 
Concretely, these are the AdS/moduli scale separation conjecture (AM-SSC)
\cite{Gautason:2018gln} and the AdS distance conjecture (ADC) 
\cite{Lust:2019zwm} (see also \cite{Alday:2019qrf}). 
Roughly, these conjectures state that the mass of certain (towers of) modes 
cannot be parametrically separated from the AdS radius.
The strong version of the ADC is reminiscent of observations 
made earlier in \cite{Gautason:2015tig} for the class of flux vacua 
without negative tension objects.
Moreover, 
we comment on the connection of our observations to the dS swampland conjecture
\cite{Obied:2018sgi,Ooguri:2018wrx,Garg:2018reu,Andriot:2018wzk,Bedroya:2019snp}.
The  evidence for these conjectures derives mostly from
the failure of contradicting  string theory constructions. 
However, concerning the no-go for dS vacua in quantum gravity, there
have also been  alternative arguments based on the concept of quantum
breaking \cite{Dvali:2014gua,Dvali:2017eba,Dvali:2018jhn,Dvali:2018fqu}.

The best understood string vacua examples are at tree-level,
e.g. flux vacua \cite{Camara:2005dc,DeWolfe:2005uu,Marchesano:2019hfb,Blumenhagen:2015kja}, 
where fairly general results can be proven.
In particular, for some tree-level type IIA flux vacua one can prove the
dS swampland conjecture \cite{Hertzberg:2007wc}, while one can more generally exclude 
dS in regimes of parametric control \cite{Junghans:2018gdb}. 
This class of flux vacua generically gives AdS vacua, supersymmetric or not, 
whose cosmological constant satisfies  the 
AM-SSC \cite{Danielsson:2018ztv}.  
However, the supersymmetric solutions were argued to fail the strong
version of the ADC \cite{Lust:2019zwm,Marchesano:2019hfb}.

Stringy AdS and dS vacua can also be constructed utilizing
not only tree-level ingredients, but also quantum, in particular
non-perturbative effects. The most famous examples are the KKLT \cite{Kachru:2003aw} 
and the large volume scenario (LVS) \cite{Balasubramanian:2005zx}. 
In both these cases, AdS minima are found in the effective 4D potential
and subsequently uplifted to dS.
We will focus on the AdS minima before the uplift mechanism. 

Taken at face  value, the KKLT  model does not satisfy the two AdS conjectures.
Since the tree-level vacua provide strong support for the conjectures, one might 
think that something is wrong or inconsistent in the quantum vacuum
construction. However, it has proven to be difficult to isolate any particular
shortcoming in the AdS minimum construction. Its full ten-dimensional description
has been analyzed in  a series of recent papers 
\cite{Moritz:2017xto,Kallosh:2018wme,
  Kallosh:2018psh,Gautason:2018gln,Hamada:2018qef,
  Carta:2019rhx,Gautason:2019jwq,Bena:2019mte} 
converging to the conclusion that
the 4D effective KKLT description captures the main aspects of this vacuum. 
The uplift to a dS vacuum is more subtle and new aspects  of
the validity of the effective field theory in the warped throat have
been investigated  in \cite{Bena:2018fqc, Blumenhagen:2019qcg}. 
So while the validity of the dS vacua is still an open question, 
the AdS vacua seem to be true counterexamples to the AdS swampland conjectures.

In this paper we take a different approach to this issue and analyze
whether the AdS swampland conjectures should rather  be modified in such a way  that these
well-established quantum AdS vacua do satisfy them. 

We will start in section \ref{sec_two} by recalling the AdS and dS  swampland
conjectures, as well as briefly describing
the emergence proposal. 
In section \ref{sec_two_b} we briefly summarize the 
manifestation of the AdS swampland conjectures for the 
well known tree-level flux compactifications. 
Here one can distinguish   DGKT-like models
\cite{DeWolfe:2005uu} with a dilute flux limit from Freund-Rubin type
compactifications, where a geometric flux becomes relevant for moduli
stabilization. Concrete examples of these two types are presented
in  appendix \ref{app_A}.

In section  \ref{sec_NPads}, we take a closer look
at the AdS vacua of KKLT and LVS. We show that for KKLT the relation
between the lightest moduli mass and the cosmological  constant
receives dominant logarithmic corrections whose origin  can be understood from a simple
scaling argument. Taking these quantum corrections seriously,  we introduce
logarithmic corrections to the initial AM-SSC. Moreover, 
in appendix \ref{app_B} we present two scenarios of how a small value
of $W_0$ can be achieved and what the implied KK scales will be.
These models suggest the presence of  $\log$-corrections in the ADC, as well. 
For the LVS we show that the two AdS swampland conjectures 
also receive $\log$-corrections. As already shown in
\cite{Conlon:2018vov}, for the AM-SSC these are subleading.

Since the swampland program forms an intricate set of intertwined statements, 
our observations should nicely align with other relevant aspects of the program. 
In section \ref{sec_five} we examine possible such connections. 
Once one accepts the presence of quantum corrections for the AdS conjectures, 
one may also expect them for the dS swampland conjecture, 
in high similarity to the $\log$-corrections
that have recently been proposed in the framework of the
trans-Planckian censorship conjecture (TCC) \cite{Bedroya:2019snp}. 
We also discuss the emergence proposal both for tree-level flux
models and non-perturbative AdS  vacua like KKLT.

\section{The swampland conjectures}
\label{sec_two}

Let us briefly review the conjectures that we will focus on, as well as the emergence proposal.

\subsection{The AdS/moduli scale separation conjecture}

The AdS/moduli scale separation conjecture (AM-SSC) \cite{Gautason:2018gln} states that 
in an AdS minimum one cannot separate the size of the AdS space 
and the mass of its lightest mode. 
Quantitatively,  the proposal is that the lightest modulus  of non-vanishing
mass has to satisfy 
\eq{
\label{mscalesep}
                           m_{\rm mod}\, R_{\rm AdS} \le c
}
where $c$ is an order one constant and $R^2_{\rm AdS}\sim
-\Lambda^{-1}$ the size of AdS.
A strong version of this conjecture says that this relation is
saturated, i.e. $m_{\rm mod}\sim R^{-1}_{\rm AdS}$.
 
A simple, enlightening example is the 5-form flux supported
$AdS_5\times S^5$ solution of the type IIB superstring. There, the
sizes $R_{\rm AdS}$ and $R_{S^5}$  of $AdS_5$ and $S^5$ are equal and both related to the 5-form flux. 
The lightest modulus %is the first Kaluza-Klein (KK) mode on $S^5$ whose 
mass scales as $m_{\rm mod}\sim R_{S^5}^{-1}$ and saturates
relation \eqref{mscalesep}.

\subsection{The AdS distance conjecture}

The AdS distance conjecture (ADC) \cite{Lust:2019zwm} states that for an 
AdS vacuum with negative cosmological
constant $\Lambda$, the limit $\Lambda\to 0$ is at infinite distance
in field space and that there will appear a tower of light states
whose masses scale as
\eq{
\label{AdSdis}
                    m_{\rm tower}= c_{\rm AdS} \, |\Lambda|^{\alpha}
}
for some constant $c_{\rm AdS}$ of order one and $\alpha>0$.
Moreover, for supersymmetric AdS vacua a stronger
version of the AdS distance conjecture was claimed, namely that in this case $\alpha=1/2$. 
Assuming that the tower of states is just the KK tower, the strong ADC 
generalizes earlier Maldacena-Nu{\~n}ez type obstructions  \cite{Gautason:2015tig} for
scale separated type II AdS flux vacua without negative tension object 
and rephrases them  as a swampland conjecture.
%In the following,
%when referring to the ADC, we shall mostly explicitly investigate 
%statement \eqref{AdSdis}, leaving the infinite distance claim open. 
%However in view of the emergence proposal (see also below), 
%there is reason to believe these two statements go hand in hand.
In the following, we shall consider the KK tower only, leaving open the possibility of 
other towers appearing.

Let us again consider ${\rm AdS}_5\times S^5$.
Having $\Lambda\sim - R_{\rm AdS}^{-2}$, we are interested in 
$R_{\rm AdS}$ becoming large. Then the
radius of $S^5$ also becomes large and the KK modes on $S^5$ scale as
\eq{
         m_{\rm KK}(S^5)\sim {1\over R}\sim |\Lambda|^{1\over 2}\,.
}  
Therefore, these KK modes constitute the tower of states for the
(strong) AdS distance conjecture with $\alpha=1/2$.

\subsection{The dS swampland conjecture}

For completeness, let us also recall the dS swampland conjecture.
In its  original  formulation \cite{Obied:2018sgi} it states that
\eq{
   |\nabla V|\ge {c\over  M_{\rm pl}} \cdot V,\,
}
where $c$ is of order one. The refined version of the conjecture \cite{Garg:2018reu,Ooguri:2018wrx}
states that either the previous inequality or 
\eq{
{\rm min}(\nabla_i\nabla_j V)\le -{c'\over  M_{\rm pl}^2} \cdot V
}
has to hold, 
where ${\rm min}(\nabla_i\nabla_j V)$ is the minimal eigenvalue of the Hessian matrix
and $c'$ is also of order one.

\subsection{Emergence of infinite distance}

In the framework of the swampland distance conjecture it has
been observed that the infinite distance emerges from integrating
out the appearing tower 
of light states \cite{Heidenreich:2017sim,Grimm:2018ohb,Heidenreich:2018kpg,Corvilain:2018lgw}.
In quantitative terms, the emergence proposal claims that the 1-loop
contribution to the moduli field metric, arising from 
integrating out a tower of states that are lighter than the natural
cut-off of the effective theory, is proportional to the tree-level metric.
Here, let us briefly recall only the main relations. 
For more details we refer to the original literature.

Say one has an effective theory in $D$ dimensions that 
has a tower of states with masses  $m_n=n \Delta m(\phi)$,
with a degeneracy of states at each mass level that scales like $n^K$.  
Note that the mass depends on the value of a modulus field $\phi$.
If $N_{\rm sp}$  of these states become lighter than the species scale
\cite{Dvali:2007wp}
\eq{\label{speciesscale}
\Lambda_{\rm sp}={\Lambda_{\rm UV} \over {N_{\rm sp}}^{\frac{1}{D-2}}},
} 
they impose a one-loop correction to the field space metric of the field $\phi$.
Here the UV cut-off $\Lambda_{\rm UV}$ is often chosen to be the Planck scale but could
in principle  be lower. 
The number of species are given by
\eq{
              N_{\rm sp}=\sum_{n=1}^{\Lambda_{\rm sp}/\Delta m}  n^K
              \approx    \left( {\Lambda_{\rm sp}\over \Delta m} \right)^{K+1}\,.
}  
The latter two relations can be inverted to give
\eq{
            \Lambda_{\rm sp}= \, ({\Lambda_{\rm UV}}) ^{D-2\over D+K-1}\:
            (\Delta m)^{K+1\over D+K-1}\,,
            \qquad
             N_{\rm sp}=\left( {\Lambda_{\rm UV}\over \Delta m} \right)^{(K+1)(D-2)\over D+K-1}\,.
}
Then the  one
loop-correction to the field space metric for the modulus
$\phi$ in $D$ dimensions can be written as
\eq{
\label{emergone}
        G^{\rm loop}_{\phi\phi}\sim  {\Lambda_{\rm sp}^{D+K-1} \over
          M_{\rm pl}^{D-2}} {  \big( \partial_\phi \Delta
                     m(\phi) \big)^2 \over \big( \Delta
                     m(\phi) \big)^{K+3}}\,.
}
In section \ref{sec_five}, from requiring $G^{\rm loop}_{\phi\phi}\sim G^{\rm tree}_{\phi\phi}$
we will determine the cut-off scale $\Lambda_{\rm sp}$. An analogous
logic was followed in \cite{Blumenhagen:2019qcg} for the effective theory in the warped throat.

\section{Tree-level vacua}
\label{sec_two_b}

Before we turn to the well-known quantum AdS vacua, we review
whether tree-level flux vacua comply with the 
aforementioned AdS conjectures\footnote{We are indebted to Daniel
  Junghans and Timm Wrase for pointing out some misconceptions in an
earlier  version of this paper.}. 

The construction of string vacua usually starts with an assumed
Ricci-flat background equipped by extra fluxes and instantons.
Then one looks for minima of the lowa energy effective action that 
stabilizes the moduli in a controlled regime. In order
to determine the KK scale one has to solve an eigenvalue problem
for fluctuations around the background. However, for that purpose 
one actually has to use the fully backreacted metric. This is often 
not possible and one hopes that a naive estimate using the initial
background plus some control arguments give already a 
good estimate\footnote{That the backreaction can be essential
for seeing some precise cancellations for models with geometric flux
was nicely demonstrated (after this paper appeared) in \cite{Font:2019uva}.}.

\subsection{Type IIA flux models}
\label{subsec_DGKT}

The best understood examples of AdS minima in string theory are 
just type IIA and type IIB flux compactifications on Calabi-Yau
manifolds. 
In type IIA one can stabilize all closed string moduli via R-R and
$H_3$ form fluxes. Classes of such concrete models have first been analyzed
in \cite{DeWolfe:2005uu} and have been called DGKT models.
More flux models of this type were considered recently in
\cite{Marchesano:2019hfb}.

A typical example for the isotropic six-torus  is presented in
appendix \ref{app_a}, where we also take into account that 
in general there exists more than just a single   KK scale.
In these models one has a dilute flux limit
that implies that the KK scales can be made parametrically larger
than the masses of the moduli. In the limit $\Lambda\to 0$ 
some fluxes have to become infinite implying that also 
some of the moduli become infinite. Therefore, $\Lambda\to 0$ 
is reached at infinite distance in field space.
As far as we can tell, all AdS flux models of this type
studied in \cite{DeWolfe:2005uu,Marchesano:2019hfb}
satisfy the relation
\eq{
                      m_{\rm mod}\sim |\Lambda|^{1\over 2}
}
between the mass of the lightest modulus and the cosmological constant. This
includes also the non-supersymmetric models.
Therefore, for DGKT models the AM-SSC
is satisfied. However, as also claimed in \cite{Lust:2019zwm},
the relevant supersymmetric DGKT vacua  do not satisfy the strong version 
of the ADC but only its weak form with 
$\alpha<1/2$, which is also satisfied for the non-supersymmetric ones.

\subsection{Geometric fluxes and Freund-Rubin models}
\label{subsec_FR}

Another well known class of AdS minima are Freund-Rubin~\cite{Freund:1980xh} backgrounds.
The standard example is the 5-form flux supported
$AdS_5\times S^5$ solution of the type IIB superstring, whose
effective description we recall in the following.

Defining $\rho=R/M_{\rm pl}$ as the radius  of the $S^5$ in Planck
units, the string scale and the 5D Planck scale are related as $M_{\rm pl}^8=M_{s}^8 \rho^5$.
Going to Einstein-frame and performing dimensional reduction of
the   10D type IIB  Einstein-Hilbert term and the kinetic term for the
5-form flux on the fluxed $S^5$,
one obtains the 5D effective potential
\eq{ 
\label{effpotads}
                   V\sim M_{\rm pl}^5 \left( -{1\over \rho^2} + {f^2\over \rho^5}   \right)\,.
}
Here $f\in \mathbb Z$ is the quantized 5-form flux
and the first term is the contribution of the internal curvature.
The AdS minimum is at $\rho_0^3= 5f^2/2 $, where the cosmological constant 
is given by $\Lambda\sim -\rho_0^{-2}M_{\rm pl}^2$.
The mass of the modulus $\rho$ can be determined as
\eq{
                      m^2_{\rho} =G^{\rho\rho} \partial^2_\rho
                     V\big\vert_0 \sim {M_{\rm pl}^2\over \rho_0^2}
}
with the metric on the moduli space $G_{\rho\rho}\sim \rho^{-2}$. 
Therefore, the mass of the $\rho$ modulus scales in the same way as
the geometric KK  scale.

This seems to be a generic feature for models where curvature
terms are relevant for moduli stabilization. In the framework of 4D flux 
compactifications this is described by turning on so-called
geometric fluxes. A typical example of this kind is presented in 
appendix \ref{app_b}. As in the example before, the $\Lambda\to 0$
limit is reached at infinite distance in field space.
In these scenarios, there is no dilute flux limit and
the KK scale is of the same order as the moduli mass scale.
The same feature appears for the non-geometric type IIB flux models 
presented in \cite{Camara:2005dc,Blumenhagen:2015kja}. 

Therefore, irrespective of supersymmetry, these models satisfy both the AM-SSC
and the strong ADC.

\subsection{Generic scaling of moduli masses}

We will now provide a simple argument why for classical flux
compactifications the moduli masses are generally expected to 
scale like $|\Lambda|^{1\over 2}$.
A generic contribution to the flux induced scalar potential
scales like $V=A \exp(- a \phi)$, where $\phi$ is a canonically normalized
modulus. Such terms balance against each other so that the
cosmological constant is expected to also behave as $\Lambda\sim -\exp(- a \phi)$.
Similarly, the masses around the minimum will be given by
\eq{
            m_{\rm mod}^2\sim \partial_\phi^2 V \sim  
            e^{-a\phi} 
}
so that $m_{\rm mod}\sim |\Lambda|^{1/2}$ is to be expected for a 
generic tree-level flux compactification. 

\section{Non-perturbative AdS vacua}
\label{sec_NPads}

In this section, we investigate the two AdS swampland conjectures for
the KKLT and the LVS. These vacua are genuinely
non-perturbative, in the sense that 
tree-level contributions are balanced against
non-perturbative effects.

\subsection{The KKLT AdS vacuum}
\label{sec_three}

Let us first consider the KKLT AdS minimum \cite{Kachru:2003aw} for the 
single K\"ahler modulus $T=\tau+i\theta$. 
Here $\tau$ measures the size of a 4-cycle and
$\theta$ is an axion. 
After stabilizing the complex structure and axio-dilaton moduli via
three-form fluxes, the effective K\"ahler- and superpotential of KKLT
is defined by  
\eq{
                 K=-3\log(T+\ov T)\,, \qquad   W=W_0 + A e^{-aT}\,.
}
Here $W_0<0$ is the value of the flux induced superpotential in its
(non-supersymmetric) minimum and the second term in $W$ arises from a
non-perturbative effect like a D3-brane instanton or gaugino
condensation on D7-branes.
The resulting scalar potential after freezing the axion reads
\eq{
\label{hannover96}
                  V_{\rm KKLT}={a A^2 \over 6\tau^2} e^{-2
                    a\tau}(3+a\tau) +{a A 
                    W_0\over 2\tau^2} e^{-a\tau}\,.
}
The supersymmetric AdS minimum of this potential is given by the 
solution of the transcendental equation
\eq{
\label{KKLTmineq}
               A(2a\tau+3)=- 3W_0 e^{a\tau}
}
and leads to a negative cosmological constant 
\eq{
\label{KKLTLam}
                 \Lambda= -{a^2 A^2 \over 6\tau} e^{-2a\tau} \,.
}
In view of the ADC, we first observe that
$\Lambda\to 0$ means $\tau\to \infty$, which is at infinite distance
in field space. 
Note that on an isotropic manifold the naive geometric  KK scale can be expressed as
\eq{
\label{KKnaiveKKLT}
                    m_{\rm KK}\sim {1\over \tau}
}
and hence is exponentially larger than any scale $|\Lambda|^{\alpha}$
expected from the ADC \eqref{AdSdis}.
 
However, one has to keep in mind that one needs an exponentially small
$W_0$. In appendix \ref{app_B}, we argue that this requires that
the background becomes highly non-isotropic so that the naive estimate
of the KK scale \eqref{KKnaiveKKLT} is not satisfied for the lightest KK
or winding modes. In fact for the toroidal example in appendix \ref{app_B1}  we
find at leading order
\eq{   
\label{KKtrueKKLTa}
m^2_{{\rm w}}\sim \tau
                   e^{-2a\tau} \,\sim \tau^2
                   |\Lambda|\sim \log^2( -\Lambda)  \, |\Lambda|\,
}
and in appendix \ref{app_B2} for the better controlled strongly warped throat 
\eq{
\label{KKtrueKKLTb}
                  m^2_{\rm KK}\sim {1\over \tau^2}\,  {e^{-{2\over 3}a\tau}\over
                \tau^{1\over 3}}\sim {1\over {\log^2( -\Lambda)}} |\Lambda|^{1\over 3}\,.
}
In both cases  the masses scale exponentially with $\tau$ and feature
$\log$-corrections.
Up to these corrections, in the toroidal case the strong ADC is
satisfied while the better controlled warped throat scenario 
only satisfies the ADC with $\alpha=1/6$.

It is also known that the effective mass of the K\"ahler modulus $\tau$ turns out to
be much smaller than the naive   KK-scale \eqref{KKnaiveKKLT}, in fact it is the lowest mass scale
in the problem.
This is then the relevant scale for the AM-SSC.
In the minimum of the potential one can determine
\eq{
\label{KKLTmodmass}
                     m_{\tau}^2 = K^{T\ov T} \partial^2_\tau
                     V\big\vert_0 =  {a^2 A^2  \over 6\tau}(2+5
                     a\tau+2a^2 \tau^2)\,
                     e^{-2a\tau}\,
}            
which indeed contains the desired factor $\exp(-2a\tau)$. 
Thus, one obtains the relation
\eq{
                     m^2_{\tau}=  - (2+5
                     a\tau+2a^2 \tau^2)\, \Lambda \,.
}
Now, neglecting $\log\log$-corrections, for  large $\tau\gg 1$ one can invert
\eqref{KKLTLam}  
\eq{
                 a\tau= -b_1 \log(-\Lambda)+b_0
}
with $b_1$ and $b_0$ positive constants of order one.
Thus, one can express $m^2_{\tau}$ as
\eq{
\label{KKLTscaling}
      m^2_{\tau}=  - \Big(c_2^2 \log^2(-\Lambda)+c_1 \log(-\Lambda) +  c_0\Big)
      \, \Lambda \,
}
with $c_2>0$.
After reintroducing powers of the Planck scale and working in the limit
$\Lambda\to 0$,  we can express the mass in the intriguing way
\eq{
\label{AdSconjKKLT}
                   m_{\tau}\sim
                   -c_2\log\big(-{\textstyle{\Lambda\over
                           M^2_{\rm pl}}}\big)  \,
            \big|\Lambda\big|^{1\over 2}\,.
}
Note that $|\Lambda|<M_{\rm pl}$ is required for the effective theory to
be controllable.  Moreover, in the limit $\Lambda\to 0$ the mass scale
still approaches zero.

Therefore, in comparison to the (classical) AM-SSC there
appears a logarithmic correction. We propose
\eq{
                           m \,R_{\rm AdS} \le c \log( R_{\rm AdS} M_{\rm pl})\,
}
to be the quantum generalization of the AM-SSC.
This is a weaker bound than the classical version \eqref{mscalesep} so
that a slight ($\log$ type) scale separation between the internal space and
the light mode is allowed. 

Similarly, as shown in \eqref{KKtrueKKLTa} and \eqref{KKtrueKKLTb} 
we also found $\log$-corrections to the ADC. 
Therefore we summarize that  for quantum vacua
like KKLT, where a non-perturbative contribution is balanced against a
tree-level one, it seems that there appears a logarithmic correction
to the result for simple perturbative vacua.

\subsection{The large volume AdS vacuum}
\label{sec_four}

Let us analyze another prominent example, namely
the large volume scenario (LVS). Recall that here one has a
swiss-cheese Calabi-Yau threefold with a large and a small K\"ahler
modulus, $\tau_b$ and $\tau_s$. The precise definition can 
be found in \cite{Balasubramanian:2005zx,Cicoli:2008va}. 
What is important here is the final form of the scalar potential
\eq{
\label{LVSpot}
           V_{\rm LVS}=\lambda { \sqrt{\tau_s} e^{-2a\tau_s}\over
             {\cal V}} -\mu  {\tau_s e^{-a\tau_s}\over {\cal
               V}^2}+{\nu \over {\cal V}^3}
}
where the total volume is ${\cal V}\approx \tau_b^{3/2}$.
Let us recall a few relevant steps from the original
paper \cite{Balasubramanian:2005zx}.
Solving the minimum condition $\partial_{\cal V} V_{\rm LVS}=0$ one
finds
\eq{
\label{condLVSa}
 {\cal V}={\mu\over \lambda}  \sqrt{\tau_s} e^{a\tau_s}
\left(1\pm \sqrt{1-{3\nu\lambda\over \mu^2 \tau_s^{3/2}}} \right) 
}
and from  $\partial_{\tau_s} V_{\rm LVS}=0$ one obtains
\eq{
\label{condLVSb}
               \left(1\pm \sqrt{1-{3\nu\lambda\over \mu^2 \tau_s^{3/2}}}
               \right) \left( {1\over 2}-2a\tau_s\right)=(1-a\tau_s)\,.
}               
Now, one proceeds by working in the perturbative regime $a\tau_s\gg 1$, in
which case the two relations can be solved analytically, yielding the
values of the moduli in the LVS minimum
\eq{
\label{LVSmin}
                       \tau^{0}_s &=\left({4\nu\lambda\over
                           \mu^2}\right)^{2\over 3}\,,\qquad
                   {\cal V}^0={\mu\over 2\lambda} \sqrt{\tau^0_s}\, e^{a \tau^0_s}\,.
}
However, plugging this back into the potential \eqref{LVSpot} one gets
zero, indicating a sort of extended no-scale structure. Therefore, to
find the actual non-vanishing value of the potential in the LVS
minimum, one has to compute the next order
in $1/\tau_s$ \cite{Blumenhagen:2009gk}. The only approximation
we did is in the solution to \eqref{condLVSb}. Thus there will be a
correction to $\tau_s^0$, which at leading order is a just a shift by
a constant $\tau^0_s\to \tau^0_s + c/a$, which one can show to be positive.
The value of the cosmological constant will then be
\eq{
                    \Lambda \sim -{3c \lambda^2 e^{-3c} \over  \mu a \:\tau_s^0}\,
                    e^{-3a\tau_s^0}  \left(1+ O({\textstyle{1\over \tau_s}})\right)\,
}
which is indeed negative. 

The lightest modulus in the game is ${\cal V}$, whose mass can be
determined by first integrating out $\tau_s$ and taking the second
derivative of the effective potential with respect to  ${\cal V}$ 
(see also \cite{Conlon:2018vov}).
After solving  $\partial_{\tau_s} V=0$, we can write the effective
potential as
\eq{
                  V_{\rm eff}({\cal V})={1\over {\cal V}^3}\left(
                    \nu +{\mu^2\over \lambda} \tau_s({\cal V})^{3\over
                      2} \Big(g({\cal V})^2-g({\cal V}) \Big)
\right)\,
}
with 
\eq{
g({\cal V})=
    2\left(\frac{1 - \phantom{4} a\tau_s({\cal V})}{1 - 4a\tau_s({\cal V})}\right)
  ={1\over 2}\left( 1 -{\textstyle {3\over 4a\tau_s({\cal V})}} +\ldots\right).
}
Here $\tau_s$ depends implicitly on ${\cal V}$.
Now, using that at leading order $\partial\tau_s/\partial {\cal
  V}\approx (a{\cal V})^{-1}$ we realize that the leading order term (in $1/\tau_s$)
again cancels \footnote{We thank Joe Conlon for pointing this out to us.} so that 
\eq{
                  m^2_{\cal V} =  K^{{\cal V}{\cal
                      V}} \partial^2_{\cal V}
                     V_{\rm eff}\big\vert_0 \sim {\lambda^2\over \mu a
                       \,\tau_s^0}  e^{-3a\tau_s^0}
 \left(1+ O({\textstyle{1\over \tau_s}})\right)\,.
}
Therefore, for the LVS AdS minimum we have found the relation
\eq{
       m^2_{\cal V}\sim    |\Lambda|\, \Big(c_0+ {c_{-1}\over \log(-\Lambda)} +\ldots\Big)\,,
}
which means that in the limit $\Lambda\to 0$ the LVS satisfies the
strong (classical) AM-SSC. However, also for LVS there will be  subleading 
$\log$-corrections. 
In contrast to KKLT, here the first two coefficients are vanishing
i.e. $c_2=c_1=0$ which presumably 
is due to the extended no-scale structure and the perturbative
stabilization of ${\cal V}$.
For LVS one can have $W_0=O(1)$ so that the naive estimate for the KK
scale is justified
\eq{
              m_{\rm KK}^2\sim {1\over {\cal V}^{4\over 3}}
              \sim    {1\over \tau_s^{2\over 3}}
              e^{-{4\over 3} a\tau_s}\sim  {1\over \log^{2\over 9}
                |\Lambda|}  |\Lambda|^{4\over 9}\,.
}
This result is similar to the KK modes \eqref{KKtrueKKLTb} for the KKLT model in the
warped throat. 
Thus, in the ADC one expects again extra quantum $\log$ corrections and $\alpha=2/9$.

\section{Other swampland conjectures}
\label{sec_five}

In this section we discuss the implications and relations of the $\log$-correction
to other swampland conjectures. First, following a similar reasoning
as in  section \ref{subsec_DGKT}, we provide a general argument
for the appearance of such corrections.

\subsection{Origin of $\log$-corrections}
 
To see the origin of the $\log$-corrections consider a typical
non-perturbative contribution to the scalar potential, which in
canonically
normalized variables takes the following double-exponential form
\eq{
\label{toypoti}
              V\sim A \,e^{-c\phi}\, e^{-(b e^{a\phi})}+ V_{\rm others}\,.
}
The other corrections can be perturbative or non-perturbative,
depending on the nature of the model. If moduli stabilization occurs
such that the first term balances the terms in $V_{\rm others}$, the
size of the first one is expected to set the scale for the potential 
and the masses in the minimum.  
Computing its second derivative with respect to $\phi$ one gets
\eq{
            m^2\sim \partial_\phi^2 V \sim 
            \Big(c^2 +2abc\, e^{a\phi}
               - b a^2 \, e^{a\phi} + (ab)^2 e^{2a\phi} \Big) \, V \, .
}
Inverting \eqref{toypoti} one can write
\eq{
              e^{a\phi}\sim -{1\over b} \log\left( {V\over A} \right)=-b_1
              \log |V| +b_0
}
so that
\eq{
      m^2 \sim  - \Big(c_2^2 \log^2(|V|)+c_1 \log(|V|) +  c_0\Big)
      \, V \,.
}
We observe that these terms take  precisely the form of  those
that we found for KKLT in \eqref{KKLTscaling}. One can well imagine
that for a full model the potential will be more complicated so that
like in LVS also further
subleading corrections $\log^{-n}(|V|)$ ($n\ge 1$) will  appear.

Thus, we conclude that the logarithmic corrections are genuinely
related to the appearance and relevance of non-perturbative effects in
the scalar potential. In the moment that such genuinely 
non-perturbative vacua exist in string theory, 
the AdS swampland conjectures are expected to receive $\log$-corrections.

\subsection{Trans-Planckian Censorship Conjecture}

If the AdS swampland conjectures receive such corrections, it is
natural to expect that also the dS swampland conjecture will be changed.
Computing the first derivative of \eqref{toypoti}, a natural
guess would be 
\eq{
\label{dSquantum}
            {|\nabla V|}\geq V  \Big( c_1\log\left({|V|}\right)+c_2\Big)\,.
}
In contrast to the AdS swampland  conjectures this relation is supposed
to hold not only at a specific point in field space (namely the minimum)
but at every  point. It remains to be seen whether such a strong local bound really makes sense. 
In any case, it is remarkable that  the right hand side
could vanish for $V=\exp(-c_2/c_1)$, thus potentially allowing dS vacua\footnote{
Utilizing quantum effects to generate stable dS vacua has been discussed in e.g. \cite{Dasgupta:2018rtp,Danielsson:2018qpa}.}.

It has been recently suggested that a more ``global'' version 
might be the more general statement.
In \cite{Bedroya:2019snp} an underlying quantum gravity reason for the dS swampland
conjecture was proposed, namely the so-called 
trans-Planckian censorship conjecture (TCC). It proposes that
sub-Planckian fluctuations must stay quantum and should
never become classical in an expanding universe with Hubble constant $H$.
More quantitatively it says
\eq{
              \int_{t_i}^{t_f} dt\, H < \log\left( {M_{\rm pl}\over H_f}\right)\,,
}            
for more details consult \cite{Bedroya:2019snp}. Two  points
are to be emphasized here, namely
that this conjecture is not local (as it involves an initial and final
time), and the appearance of a logarithm on the right hand side.
For a monotonically  decreasing positive potential, the authors of
 \cite{Bedroya:2019snp} derived from the TCC a global version of the 
dS swampland conjecture
\eq{
\label{TCCdS}
        \left\langle{-V'\over
            V}\right\rangle\bigg\vert_{\phi_i}^{\phi_f}>{1\over \Delta \phi}
\log\Big( {V_i\over M}\Big)  +{2\over \sqrt{(d-1)(d-2)}}
}
where $\left\langle{-V'\over
            V}\right\rangle\Big\vert_{\phi_i}^{\phi_f}$ denotes the
        average of $-V'/V$ in the interval $[\phi_i,\phi_f]$.
Here $V<M<M_{\rm pl}$ and $M$ is a  mass scale that can be lower than
the Planck-scale.

Let us check that a potential of the generic form
\eq{
               V(\phi)=A e^{-c\phi} \,  e^{-(b e^{a\phi})}
}
satisfies this averaged dS swampland conjecture.
Note that for $a,b,c>0$ this potential is indeed positive and 
monotonically decreasing.
For the average value we can directly compute
\eq{
\left\langle{-V'\over
            V}\right\rangle\bigg\vert_{\phi_i}^{\phi_f}&={1\over \Delta
          \phi}   \int_{\phi_i}^{\phi_f} d\phi \left(c + ab\, e^{a\phi}\right)
              = c +{b\over \Delta \phi} \left(
                e^{a\phi_f}-e^{a\phi_i}\right)\\
                     &> c   - {b\over \Delta \phi} e^{a\phi_i}> 
                        c+{1\over \Delta \phi} \log\left({V_i\over A}\right)\,.
}
This has precisely the form \eqref{TCCdS} so that we can state that 
non-perturbative contributions to the scalar potential induce 
the $\log$-corrections in the TCC
derived dS swampland conjecture \eqref{TCCdS}. 
Moreover, we observe that for the three terms in the KKLT potential 
\eqref{hannover96}, one gets the parameters $c\in\{\sqrt{8/3},\sqrt{2/3}\}$ 
which both satisfy  $c\ge \sqrt{2/3}$, the value  appearing in  \eqref{TCCdS}. 

We consider this connection to the TCC as further evidence  
for  the appearance of $\log$-corrections in the various
swampland conjectures.

\subsection{Emergence for KKLT}

Finally, we  comment on the emergence proposal. 
Before we come to the KKLT model let us first  consider 
tree flux compactifications.

For compactification on $S^5$ one needs to take heed of the degeneracy
of KK modes of mass $m_n=n \Delta m = n M_{\rm pl}/\rho$. 
This is given by the dimensionality of the space of harmonic functions of 
homogeneous degree $n$, which for the 5-sphere goes as $n^4$. 
Applying our general result \eqref{emergone} for the one-loop correction to
the field space metric and setting it equal to the tree-level metric
$G_{\phi\phi}^{\rm tree}\sim \rho^{-2}$  we obtain  $\Lambda^8_{\rm
  sp}=M^3_{\rm pl} (\Delta m)^5= M^8_{\rm pl}/\rho^5$. Taking the
relation
between the string scale and the $D$ dimensional Planck scale into
account it follows $\Lambda_{\rm sp}\sim M_s$. We expect that this
relation will appear for all tree-level  flux compactifications so that
the cut-off of these models is simply the string scale.

Next we will analyze the implications of the emergence proposal in the KKLT setting, 
both for the toroidal model and the strongly warped throat.

\subsubsection*{Emergence for toroidal  KKLT}

In appendix \ref{app_B1} we found a (non-degenerate) tower of states with discretized
masses scaling as \eqref{KKtrueKKLTa}.
At leading order in $\tau$, the 1-loop correction \eqref{emergone}
to the field space metric of the modulus $\tau$ reads
\eq{
          G_{\tau\tau}^{\rm loop}\sim {\Lambda_{\rm sp}^3 \over M_{\rm
              pl}^3} {a^2\over \sqrt{a\tau}} e^{a\tau}\,.
}
Imposing that this is proportional to the tree-level metric
$G_{\tau\tau}^{\rm tree}\sim \tau^{-2}$, one can determine the value
of the cut-off of the effective theory as
\eq{
\label{KKLTtoruscutoff}
             \Lambda^3_{\rm sp}\sim {e^{-a\tau} \over \tau^{3\over 2}}
             M_{\rm pl}^3\,.
}
This is reminiscent of the dynamically generated mass scale
$\Lambda_{\rm SQCD}$ of the SYM theory that undergoes gaugino
condensation. This scale is usually given by 
${\Lambda^3_{\rm SQCD}= e^{-a/g^2}   M^3}$, 
where $M$ denotes a UV cut-off scale.
Noting that $g^{-2}\sim \tau$ we can write the KKLT cut-off as
\eq{
             \Lambda^3_{\rm sp}\sim 
             e^{-{a\over g^2}}  (g\, M_{\rm pl})^3\sim \Lambda^3_{\rm SQCD}\,.
}
Thus the cut-off of the KKLT model is the scale at which the
implicitly assumed gaugino condensation of the confining gauge theory occurs, 
while the UV cut-off of the gauge theory itself is not simply the Planck
scale but rather $M\sim \Lambda_{\rm UV}\sim gM_{\rm pl}$ as suggested by the weak gravity conjecture.

\subsubsection*{Emergence for warped throat KKLT}

Next we analyze our second example from appendix
\ref{app_B2}, where the small value of $W_0$ is generated by a
strongly warped throat. As shown in \cite{Blumenhagen:2019qcg}, in this case there exists a
tower of highly red-shifted KK modes localized at the tip of the
throat with masses 
\eq{
       \Delta m_{\rm KK}\sim { |Z|^{1\over 3}\over \tau^{1\over 2}
         y_{\rm UV}}\,,
} 
where $Z$ denotes the conifold (complex structure) modulus and $y_{\rm
  UV}$ is the length of the KS throat before it reaches the bulk
Calabi-Yau. It was argued in \cite{Blumenhagen:2019qcg} that these KK modes are lighter
than the cut-off of the effective theory and that their one-loop
contribution corrects the second (subleading) term in the K\"ahler potential
\eq{
              K=-3\log(T+\ov T) +  c  {|Z|^{2\over 3}\over (T+\ov T)}\,.
}
Using the general relation \eqref{emergone} and setting this one-loop
correction equal to the K\"ahler metric $G_{\tau\ov \tau}$ (second
term) one finds
for the species scale
\eq{
                 \Lambda_{\rm sp}^3\sim {|Z|\over \tau^{3\over 2} y_{\rm
                     UV}} M_{\rm pl}^3\,.
} 
Since the first term in the above K\"ahler potential is
also present in the unwarped case, we expect it to 
emerge from integrating out the tower of heavier bulk KK modes
\eqref{KKnaiveKKLT} with mass scale $\Delta m_{\rm KK,h}\sim 1/\tau$.

In the limit where  the throat just fits into the warped Calabi-Yau volume, one
can determine  the cut-off $y_{\rm UV}$  as
\eq{
                   y_{\rm UV}\sim -\log\left( {|Z|\over \tau^{3\over 2}}\right)\,.
}
Now we stabilize the K\"ahler modulus $\tau$ via KKLT which gives the
relations
\eq{
             |Z|\sim |W_0|\sim \tau e^{-a\tau}\,,\qquad   y_{\rm
               UV}\sim \tau
}
so that the species scale can be expressed as 
\eq{
             \Lambda_{\rm sp}^3\sim  {e^{-a\tau} \over \tau^{3\over 2}}
             M_{\rm pl}^3\,,
}
which is the same result \eqref{KKLTtoruscutoff} as for the toroidal
setting. Therefore,  also for the warped throat KKLT model the relation 
to  $\Lambda_{\rm SQCD}$ holds.

\section{Conclusions}

In this paper, we have investigated the behaviour of the KKLT and LVS constructions
with regard to the AdS scale separation and distance conjectures.
To this end, we have identified the relevant towers of light states and provided
two concrete examples of realizing an exponentially small mass scale for the KKLT model.
Driven by confidence in the consistency of the aforementioned AdS vacua,
we propose $\log$-corrections to the tree-level AdS swampland conjectures.
Extending our reasoning, we expect similar $\log$-corrections to the dS swampland conjecture.
These might be in the same spirit as the $\log$-corrections that were found 
for the ``average" of the dS swampland conjecture in the recently proposed TCC.
Whether a stronger, local version of a quantum dS swampland conjecture
could make sense is left for future analysis.

Additionally, we analyzed the consequences 
of imposing the emergence proposal. For tree-level flux
compactifications we found that the cut-off scale is simply the string
scale. For the KKLT model, it is remarkable that both proposed 
scenarios for generating an exponentially small $W_0$ lead to a cut-off scale
reminiscent of the dynamically
generated scale for the condensing SYM theory. 
It is certainly encouraging that our observations seem to fit well 
within the broader swampland picture. 

%%%%%%%%%%%%%%%%%%%%%%%%%%%%%%%%%%%%%%%%%%%%%%%

\vspace{0.3cm}
\subsubsection*{Acknowledgements:}
We thank Joe Conlon, Daniel Junghans, Dieter L\"ust, Eran Palti, Lorenz Schlechter and Timm Wrase
for very helpful discussions and 
for useful comments about an earlier version of the paper.

\newpage

\appendix

\section{KK scales in type II flux compactifications}
\label{app_A}

In this appendix we present two simple though typical type IIA flux
compactifications on the isotropic $T^6$. The first one only contains
R-R and $H_3$-form fluxes and as expected features a dilute flux limit that
allows to separate the KK scale from the moduli mass scale.
The second one also contains geometric fluxes, in which case there
will be no dilute flux limit and the moduli masses are of the same
scale as the KK modes. Freund-Rubin type compactifications are
fully fledged 10D uplifts of such effective models.

\subsection{A typical DGKT model}
\label{app_a}

Let us consider type IIA orientifolds with fluxes on an isotropic
six-torus. Here one has three chiral superfields $\{S,T,U\}$ whose
real parts are defined as
\eq{
          \tau=r_1 r_2\,,\qquad   s=e^{-\phi} r_1^3\,,\qquad
          u=e^{-\phi} r_1 r_2^2\,.
}
The axions do not play any role in the following and will in all
examples be stabilized at vanishing value.
The K\"ahler potential is given as
\eq{
       K=-3 \log(T+\ov T)-3 \log(U+\ov U)-\log(S+\ov S)\,.
}
Now we turn on just R-R fluxes and $H_3$-form flux so that the flux induced
superpotential reads
\eq{
           W=if_0 T^3 - 3if_4 T + ih_0 S + 3ih_1 U\,.
}
Then there exist both supersymmetric and non-supersymmetric AdS minima. 
For instance in the supersymmetric vacuum, the saxionic moduli are stabilized at
\eq{
                 \tau=\kappa {f_4^{1\over 2}\over f_0^{1\over 2}}\,,\qquad
                  s={2\kappa\over 3} {f_4^{3\over 2}\over f_0^{1\over 2} h_0}\,,\qquad
                  u={2\kappa} {f_4^{3\over 2}\over f_0^{1\over 2} h_1}\,
}
with $\kappa=\sqrt{5/3}$. For the non-supersymmetric minima only the
numerical prefactors change.
The  effective masses of the moduli all scale in same way as
\eq{
                  m_{\rm mod}^2\sim -\Lambda\sim {f_0^{5\over 2} h_0 h_1^3\over
                    f_4^{9\over 2}} M_{\rm pl}^2\,.
}
Therefore the model satisfies the AM-SSC.
The two KK scales are
\eq{
                m_{{\rm KK},1}^2 &= {M_s^2\over r_1^2}={M_{\rm pl}^2\over s\,\tau\,u}={f_0^{3\over 2} h_0 h_1\over
                    f_4^{7\over 2}} M_{\rm pl}^2 \,, \\
               m_{{\rm KK},2}^2 &= {M_s^2\over r_2^2}={M_{\rm pl}^2\over \tau\,u^2}={f_0^{3\over 2}  h_1^2\over
                    f_4^{7\over 2}} M_{\rm pl}^2     \,.        
} 
To be in the perturbative regime ($g_s\ll 1$) we now choose $f_0, h_0,h_1=O(1)$ and
$f_4 \gg 1$. 
In this regime the KK scales are parametrically larger than the moduli masses  and one
has
\eq{         
          m_{{\rm KK},i}^2\sim |\Lambda|^{7\over 9}\,,
}
thus satisfying the ADC with $\alpha=7/18$, both in
the supersymmetric and non-supersym\-metric case.  This reflects the
fact
that type IIA  flux compactifications admit a dilute flux limit (where $f_4\to\infty$).                 

\subsection{A type IIA  model with geometric flux}
\label{app_b}

The Freund-Rubin background $AdS_5\times S^5$ is an example 
of a flux compactification, where for moduli stabilization the
curvature of the internal space is essential (as seen in e.g. the effective 
potential eq. \eqref{effpotads}). Such backgrounds can be  described 
by turning on geometric flux $\omega$ in the effective theory. 

As a typical simple model we consider the superpotential 
\eq{
           W=f_6 + 3f_2 T^2 - \omega_0 S T -3 \omega_1  U T\,.
}
Then the saxions are stabilized in a supersymmetric AdS minimum at
\eq{
                 \tau={1\over 3} {f_6^{1\over 2}\over f_2^{1\over 2}}\,,\qquad
                  s=2{f_2^{1\over 2}  f_6^{1\over 2}\over  \omega_0}\,,\qquad
                  u=2{f_2^{1\over 2}  f_6^{1\over 2} \over \omega_1}\,
}
and receive masses that  scale as
\eq{
                  m_{\rm mod}^2\sim -\Lambda\sim { \omega_0\, \omega_1^3\over
                    f_2^{1\over 2}  f_6^{3\over 2}   } M_{\rm pl}^2\,.
}
In this case the two KK scales are
\eq{
       m_{{\rm KK},1}^2 = {  \omega_0 \,\omega_1\over
                     f_2^{1\over 2}  f_6^{3\over 2} } M_{\rm
                     pl}^2\,,\qquad
       m_{{\rm KK},2}^2 = {  \omega_1^2\over
                     f_2^{1\over 2}  f_6^{3\over 2} } M_{\rm
                     pl}^2
}
which satisfy $m_{{\rm KK},1}^2 \sim m_{\rm
  mod}^2/\omega_1^2$ and $m_{{\rm KK},2}^2 \sim m_{\rm
  mod}^2/(\omega_1 \omega_2)$. It was shown in \cite{Font:2019uva} that
taking the backreaction of the fluxes onto the metric into account, 
the geometric fluxes in the denominator also cancel and parametrically one
indeed finds $m_{{\rm KK}}^2\sim  m_{\rm mod}^2$.
Therefore, in such models with geometric flux there is no
parametric separation of the KK scale and the moduli masses, the same behavior
that occurs for Freud-Rubin compactifications. Here both the AM-SSC and the 
strong ADC are satisfied.

\section{KK scales for KKLT}
\label{app_B}

The ultimate question is what happens with the KK scale in the KKLT
scenario.
As we have seen, the naive KK scale  \eqref{KKnaiveKKLT} is not exponentially small in the 
K\"ahler modulus $\tau$, so that there seems to be no way that the
ADC can hold.
However, we have to keep in mind that for the KKLT scenario to work
one needs an exponentially small value of $W_0$ after stabilizing the complex structure moduli and the dilaton
\`a la GKP~\cite{Giddings:2001yu}. 

\subsection{A toroidal example}
\label{app_B1}

Let us consider again a simple toroidal type IIB model, 
for which we can easily compute the KK scales directly.
In this case, the real parts of the chiral superfields $\{S,T,U\}$ are defined as
\eq{
          \tau= e^{-\phi}r_1^2 r_2^2\,,\qquad   
          s=e^{-\phi} \,,\qquad 
          u=r_2/r_1\,.
}
We turn on $F_3$ and $H_3$ form flux such that the superpotential is
\eq{
      W= i f U + i h S U^2\,
}
with $f,h$ positive.
This freezes the axions completely and the saxions have to satisfy
\eq{
         u s = {f\over h}\,,
}
leading to the value $W_0=2if u$ of the superpotential along the minimum.
Therefore, the superpotential becomes very small for $u\ll 1$ while
the dilaton stays in the perturbative regime, i.e. $e^{\phi}\ll 1$.
The KK and winding scales can be computed in terms of $u$ and $\tau$ as
\eq{
                   m^2_{{\rm KK},\pm}= {M_{\rm pl}^2 \over \tau^2} u^{\pm
                     1}\,, \qquad 
                   m^2_{{\rm w},\pm}= {M_{\rm pl}^2 \over s\tau} u^{\pm
                     1}\,.
}
In the  regime of interest $s\sim u^{-1}$, the large radius regime, where the KK scale is smaller than the
winding scale, is given by $\tau u \gg 1$. As long as $\tau$ is not
stabilized this can always be satisfied by choosing $\tau$ large enough.
However, in KKLT $\tau$ is fixed as $W_0 \sim u \sim \tau
\exp(-a\tau)$, which implies $\tau u \ll 1$. Therefore,  the lightest
tower of states in KKLT is given by the winding modes 
\eq{
             m^2_{{\rm w},+}
             \sim {M_{\rm pl}^2 \over \tau} u^2
             \sim {M_{\rm pl}^2 \over \tau} |W_0|^2\,.
}
In the KKLT minimum, using \eqref{KKLTmineq} and \eqref{KKLTLam}, the winding  mass becomes
\eq{
                   m^2_{{\rm w},+}&\sim 
                   \frac{e^{-2a\tau}}{\tau} (4 a^2\tau^2+12a\tau+9) \,M_{\rm pl}^2 \\
                   &\sim  (\log^2(-\Lambda)-6\log(-\Lambda)+9)|\Lambda|
                  \,,
}
which is of the same form as \eqref{KKLTscaling}.
This satisfies the strong ADC up to $\log$ corrections. 

Thus we have seen that, if we want to have an
exponentially small value of $W_0$, the torus becomes highly
non-isotropic  and  towers of states  become  exponentially light 
$m^2_{{\rm w},+}  \sim M_{\rm pl}^2 \, \tau^{-1} \exp\left({-\sqrt{8/3}\, \phi_u}\right) $
in the canonically normalized field $\phi_u=-\sqrt{3/2} \log u$
corresponding to the complex structure modulus $u$.
This is nothing else than the tower of states that must become light
in the large $\phi_u$ regime due to the swampland distance conjecture.

One relevant concern is that for such large excursions
in the complex structure $u$, the effective theory that we used for
computing complex structure moduli stabilization is not under control
anymore. 
For instance, as we have seen we do not satisfy $\tau u\gg 1$ so that
one  radius of the torus becomes
significantly smaller than the string length.  As a consequence,  higher 
derivative terms in the effective action  might not be negligible.
Moreover,  the mass scale of these  winding  modes is lighter than the mass 
$m_{u,s}^2\sim \tau^{-3}$ of the stabilized complex structure moduli. 
Therefore, this simple toroidal model can certainly not serve as a completely
convincing flux GKP compactification with $W_0\ll 1$. Nevertheless, it
exhibits an important feature, namely that $W_0\ll 1$ goes along with
the occurrence of a tower of states whose mass scales as $m \sim \exp(-a\tau)$.

\subsection{The warped throat}
\label{app_B2}

Another option was proposed in \cite{Blumenhagen:2019qcg}, namely that a superpotential 
involving the complex structure modulus $Z$ governing the appearance of a conifold
singularity can also generate an exponentially small value for $W_0$. If $|Z|\ll
1$ the three-cycle of the conifold  becomes very small  and locally the
geometry is described by a Klebanov-Strassler(KS) throat.
For our purpose we only need a couple of relations.
First the superpotential in the minimum is given by  
\eq{
|W_0|\sim |Z|\sim \exp\left( -{2\pi  h\over g_s f}\right)
}
where $f,h$ are $F_3$ and $H_3$ fluxes supporting the strongly warped
KS throat. It was shown in \cite{Blumenhagen:2019qcg} that there exists a tower of light KK modes
localized close to the tip of the conifold with masses
\eq{
                m^2_{\rm KK}\sim {1\over y_{\rm UV}^2}\left({|Z|\over {\cal
                    V}}\right)^{2\over 3}  M_{\rm pl}^2 
}
where ${\cal V}=\tau^{3/2}$ denotes the warped volume of the
threefold. (Note that one must have ${\cal V} |Z|^2\ll 1$ in the
strongly warped throat.) Moreover, $y_{\rm UV}$ denotes the length of
the warped KS throat before it goes over to the bulk Calabi-Yau
manifold.  In the limit that the throat just fits into the Calabi-Yau
volume one can relate $y_{\rm UV}$ to the other quantities as
\eq{
                         y_{\rm UV}\sim -\log\left( {|Z|\over {\cal V}}\right)
}
(see \cite{Blumenhagen:2019qcg} for further details).
It was also found that the mass scale of these KK modes is
of the same order as the mass of the complex structure $Z$. Thus, one is
(still) at
the  limit of control of the utilized effective theory. In that respect, this
scenario is  better controlled than the toroidal model discussed before.

Now  using again  the KKLT minimum condition \eqref{KKLTmineq} 
we get\footnote{A more accurate approximation would be $ y_{\rm
    UV}\sim a\tau+\frac{1}{2}\log\tau$ but this leads to
  $\log\log |\Lambda|$ terms.}
%but this just induces an
%  overall factor $a^{-2}$ in the KK masses that is irrelevant for our
%  purposes in this paper.} 
$ y_{\rm UV}\sim a\tau$ and 
can express this exponentially small KK scale as
\eq{
            m^2_{\rm KK}&\sim  {|W_0|^{2\over 3}\over
              a^2\tau^3} M_{\rm pl}^2
              \sim \left({(2a\tau+3)^{2}\over a^8\tau^{8}}\right)^{1\over 3} 
              \,\left({ a^2 e^{-{2}a\tau}\over
                \tau}\right)^{1\over 3} M_{\rm pl}^2
                \\[3pt]
                &\sim \left({1\over \log^2( -\Lambda)} 
                - {2\over \log^3( -\Lambda)} + ...
                \right)|\Lambda|^{1\over 3}  M_{\rm pl}^2\,.
}
Up to the $\log$-term this satisfies the ADC with  $\alpha=1/6$.

We note that  ${\cal V} |Z|^2\sim \tau^{7/2} \exp(-2a\tau)$ which for
large $\tau$ is indeed much smaller than one. Therefore, stabilizing
the K\"ahler modulus via KKLT is self-consistent with using the
effective theory in the warped throat. In this respect, it also
behaves better than the toroidal model from the previous subsection.

%%%%%%%%%%%%%%%%%%%%%%%%%%%%%%%%%%%%%%%%%%%%%%%
%%%%%%%%%%%%%%%%%%%%%%%%%%%%%%%%%%%%%%%%%%%%%%%
\newpage

\bibliography{references}  
\bibliographystyle{utphys}

\end{document}